\documentclass[runningheads]{llncs}
\usepackage[T1]{fontenc}
\usepackage{graphicx}
\usepackage{booktabs}
\PassOptionsToPackage{hyphens}{url}\usepackage{hyperref}

\begin{document}
\title{Evaluating Data Quality Tools: Measurement Capabilities and LLM Integration}
\titlerunning{Evaluating Data Quality Tools}
%
\author{Tobias Rehberger\inst{1} \and
Thomas H{\"u}tter\inst{1}\orcidID{0000-0002-7190-6825} \and
Lisa Ehrlinger\inst{2}\orcidID{0000-0002-1825-0097} \and
Wolfram W{\"o}{\ss}\inst{3}\orcidID{0009-0006-8352-0205}}
\authorrunning{Rehberger et al.}
%
\institute{
Software Competence Center Hagenberg, Hagenberg, Austria\\
\email{\{tobias.rehberger,thomas.huetter\}@scch.at} \and
Hasso Plattner Institute, University of Potsdam, Potsdam, Germany\\
\email{lisa.ehrlinger@hpi.de} \and
Institute for Application-Oriented Knowledge Processing, Johannes Kepler University Linz, Linz, Austria\\
\email{wolfram.woess@jku.at}}
\maketitle              
\begin{abstract}
    High data quality is critical for reliable analytics and operational efficiency. A growing ecosystem of tools has emerged to support data quality management, ranging from lightweight open-source libraries to comprehensive enterprise platforms. This paper evaluates six data quality tools: Great Expectations, Deequ, Evidently, Informatica, Experian, and Ataccama. The evaluation criteria cover rule definition, duplicate detection, metric aggregation, and uncertainty handling, and were derived from real-world use cases of company partners. We further examine to what extent these tools integrate Large Language Models (LLMs). Our findings show that proprietary tools offer more comprehensive measurement features and emerging LLM-based assistance, while open-source tools provide flexibility at the cost of higher implementation effort. Across all tools, LLM integration remains limited to rule creation workflows. Direct data validation through LLMs is not yet supported by any of the evaluated tools.

\end{abstract}
\section{Introduction}
High-quality data is a fundamental requirement for reliable analytics, AI applications, and business decision-making. Poor data quality leads to costly decisions, operational inefficiencies, and a loss of trust among stakeholders~\cite{batini2016dataquality,Nagle_2020,redman2018impact}. As organizations increasingly rely on data to drive strategic decisions and train machine learning models, ensuring data quality has become one of the greatest challenges in corporate data management~\cite{RestatKS23,wang1996beyondaccuracy}.

To address these challenges, a growing ecosystem of DQ tools has been developed, offering functions such as profiling, validation, cleansing, and monitoring~\cite{Ehrlinger_2022,gartner2021magicquadrant,Papastergios_2025,RestatDKS25}. Recent advances in generative AI and Large Language Models (LLMs) promise to further automate and enhance DQ measurement, classification, and cleaning~\cite{ataccama-ai,BoecklingB25,glock2025}. Yet, the diversity of available tools, ranging from lightweight open-source libraries to comprehensive enterprise platforms, poses a challenge for practitioners to select a tool that fits their requirements.

This paper evaluates the capabilities of six leading DQ tools. The evaluation is structured around criteria that are derived from practical scenarios of our company partners. We examine the support for rule definition, duplicate detection~\cite{Heinrich_2025}, metric aggregation, and uncertainty handling, and investigate to what extent tools integrate LLMs. By evaluating both open-source and proprietary tools, we aim to offer actionable insights for organizations that seek suitable DQ solutions.

This paper is structured as follows. Section~\ref{sec:related} discusses related work. Section~\ref{sec:tools} describes the tool selection process and the investigated tools. Section~\ref{sec:criteria} details the evaluation criteria. Section~\ref{sec:results} presents the results and the paper concludes in Section~\ref{sec:conclusion}.

\section{Related Work}
\label{sec:related}
Several surveys have examined DQ tool functionalities. Early work by Barateiro et al.~\cite{Barateiro_2005} and Pushkarev et al.~\cite{Pushkarev_2010} compared general functionalities such as data source support, metadata repositories, and graphical interfaces, but did not investigate low-level rule capabilities or LLM integration. Ehrlinger and Wöß~\cite{Ehrlinger_2022} conducted a large-scale survey of 13 tools, covering DQ metrics and dimensions from a user perspective. Papastergios et al.~\cite{Papastergios_2025} further examined how DQ dimensions are implemented in practice across a broad set of tools. 
In contrast to these surveys, this paper focuses on the integration of LLMs, which has not been addressed in prior work.

\section{Evaluated Data Quality Tools}
\label{sec:tools}
The proprietary DQ tools included in this evaluation were selected based on the ``Magic Quadrant for Augmented Data Quality Solutions'' by Gartner~\cite{gartner2021magicquadrant}, a recognized industry standard for evaluating leading providers in the field of data-driven solutions to quality management. 
Specifically, Informatica, Experian, and Ataccama are positioned as leaders in this report, indicating both completeness of vision and ability to execute. To complement this selection with open-source perspectives, we additionally include Deequ, Great Expectations, and Evidently, which were selected based on their wide industry adoption and large user base~\cite{Papastergios_2025}. 
The evaluation of all tools was conducted using publicly available information, including official vendor websites, publicly accessible documentation, and, where applicable, GitHub repositories. 

\subsection{Great Expectations (1.4.2)}
Great Expectations\footnote{\url{https://greatexpectations.io}} can be used either as a SaaS (Software as a Service) solution (GX Cloud) or as a Python library (GX Core). It allows the definition and execution of ``expectations'', which are verifiable rules about data structure and content. 
Great Expectations supports data sources like Pandas, SQL databases, and Apache Spark, and integrates with orchestration tools such as Airflow or dbt (Data Build Tool). 
Notable features include human-readable rule definitions, auto-generated HTML reports (Data Docs), and a library of predefined checks~\cite{gx-main,gx-blog}.

\subsection{Deequ (2.0.10)}
Deequ\footnote{\url{https://github.com/awslabs/deequ}} is a library built on Apache Spark that enables the definition of ``unit tests for data'' to systematically verify DQ in large datasets~\cite{SchelterBLRSSBT19}. 
Deequ allows assumptions about the structure and properties of data to be explicitly defined and automatically checked. These assumptions are formulated as tests and can be used to detect, isolate, and correct faulty data before it enters production systems or analytical processes.
The core component is the \texttt{Verification Suite}, which defines checks. Checks are rules that describe specific quality requirements for the data, e.g., no null values or value ranges. Deequ is particularly suitable for scalable data processing workflows and enables programmatic and reproducible monitoring of DQ~\cite{deequ-medium,deequ-github}.

\subsection{Evidently (v0.7.3)}
Evidently\footnote{\url{https://www.evidentlyai.com}} is an open-source Python library designed for monitoring and evaluating DQ and data drift, particularly in machine learning (ML) applications. It enables continuous analysis of datasets and model outputs to detect deviations, quality issues, and potential risks early.
The tool provides user-friendly visualizations and interactive reports for data profiling, validation, and drift analysis. It integrates seamlessly into Python workflows and CI/CD (Continuous Integration, Continuous Deployment) pipelines. Evidently is suitable for monitoring ML systems and their data quality during operation. Additionally, it offers a complete toolkit for AI testing and monitoring via its cloud platform~\cite{evidently-docs,evidently-github}.

\subsection{Informatica}
The DQ module of Informatica\footnote{\url{https://www.informatica.com/products/data-quality/informatica-data-quality.html/en.html}} is a comprehensive and scalable solution for analyzing, cleansing, validating, and standardizing data. It offers features such as data profiling, rule-based validation, enrichment, and continuous monitoring of DQ. The module is tightly integrated into the Informatica Data Management Platform, supporting effective governance of DQ standards.

Informatica uses CLAIRE, an AI engine, to detect anomalies based on data profiling results and to suggest rules related to completeness, uniqueness, and validity. These suggestions are not generated through interactive LLM prompts but are based on existing profiling outcomes~\cite{informatica-main,claire-ai,claire-gpt}.

Rules in Informatica are defined using logical operations and can be grouped into rule sets. These rule sets allow for complex combinations using logical operators. The platform also supports deduplication with configurable similarity thresholds and confidence levels. Aggregation of rule results is possible via scorecards, which summarize metrics like average scores by dimension and rule occurrences~\cite{informatica-dedup,informatica-rules,informatica-scorecards}.

\subsection{Experian}
Experian\footnote{\url{https://www.edq.com}} offers a comprehensive suite of DQ solutions that help organizations to manage and optimize their data. Key products include Experian Aperture Data Studio, Governance Studio, and Batch tools, which support validation, cleansing, deduplication, and enrichment of data from various sources.

Experian provides built-in functionality for rule definition and validation workflows. These include native functions like \texttt{IsEmpty} and \texttt{IsNull}, and regular expression matching and custom functions that allow combining multiple rules. The platform also supports duplicate detection using advanced matching algorithms across fields such as name, address, email, and phone number~\cite{experian-main,experian-docs,experian-dedup}.

Experian enables rule grouping with configurable thresholds and optional weighting, allowing aggregation of results at the group level. However, the documentation does not specify the exact aggregation method. For uncertainty, Experian provides a categorical match status with levels such as Exact, Close, Probable, Possible, and None, indicating the confidence in duplicate detection~\cite{experian-match}.

\subsection{Ataccama (15.3.0)}
Ataccama\footnote{\url{https://www.ataccama.com/platform/data-quality}} offers a comprehensive solution to ensure and improve DQ through the Ataccama ONE platform. This suite integrates data management and governance with tools for metadata management and MDM (Master Data Management). Ataccama ONE provides features for profiling, validation, and standardization of data.

The platform uses AI-based analysis to identify data issues and automatically suggest fixes, helping to create a consistent and error-free data foundation. Ataccama supports rule creation, including technical and business rules, and allows logical combinations of rules. It also provides matching configurations for deduplication and similarity detection across records.
AI features in Ataccama ONE encompass rule generation, SQL query assistance, and catalog item descriptions. These features are embedded in workflows and support interactive rule creation based on natural language input~\cite{ataccama-blog,ataccama-docs,ataccama-ai}.
\section{Evaluation Criteria}
\label{sec:criteria}

This section introduces the criteria used to evaluate open-source and proprietary DQ tools. The criteria cover fundamental aspects of DQ management, including rule definition, duplicate detection, metric aggregation, and uncertainty handling, while also addressing the integration of LLMs for rule creation and validation. They were derived from real-world use cases of our company partners, including the continuous validation of customer master data, deduplication of contact records, and aggregation of DQ metrics across distributed data sources.

The evaluation of DQ tools is based on six criteria. The term ``rules'' refers to a formalized condition or constraint that specifies how data should conform to defined standards. Achieving Level 2 automatically satisfies the requirements of Level 1, as higher levels build cumulatively on the preceding ones. The first two criteria address LLM integration, while the remaining four cover core DQ measurement capabilities.

Before detailing each criterion, Table~\ref{tab:criteria_levels} provides a compact overview of all criteria and their applicable levels.

\begin{table}[ht]
    \centering
    \caption{Criteria Fulfillment Levels}
    \begin{tabular}{l@{\hspace{15pt}}l@{\hspace{15pt}}c@{\hspace{10pt}}c@{\hspace{10pt}}c@{\hspace{10pt}}c}
        \toprule
        \textbf{ID} & \textbf{Criteria} & \multicolumn{4}{c}{\textbf{Levels}} \\
        \midrule
        0 & Open Source & No & Yes & & \\
        1 & Quality Check with LLMs & No & Yes & & \\
        2 & Rule Creation with LLMs & - & + & ++ & +++ \\
        3 & Rule Definition (Correctness) & - & + & ++ & +++ \\
        4 & Minimality & - & + & ++ & +++ \\
        5 & Aggregation of Metrics & - & + & ++ & +++ \\
        6 & Uncertainty & - & + & ++ & +++ \\
        \bottomrule
    \end{tabular}
    \label{tab:criteria_levels}
\end{table}

\begin{definition}[Quality Checks with LLMs]
Are LLMs used to check data? (Yes/No) For example, data is passed to an LLM along with a prompt containing validation details. The result is generated according to the prompt definition.
\end{definition}

\begin{definition}[Rule Creation with LLMs]
To what extent are LLMs used for rule creation?
    \begin{description}
      \item[Level 1:] Technical rules can be generated through interaction with the LLM. For example, an LLM creates a rule with the desired functionality without writing actual code.
      \item[Level 2:] Business rules involving domain knowledge can be generated. For example, based on domain knowledge, an LLM creates a rule with the desired functionality without writing actual code.
      \item[Level 3:] The LLM automatically suggests suitable rules without user input. For example, an LLM provides rules with the desired functionality without requiring any user input.
    \end{description}
\end{definition}

\begin{definition}[Rule Definition]
To what extent can rules be defined to check a certain dimension of data quality (e.g., correctness)?
    \begin{description}
      \item[Level 1:] Simple technical rules, e.g., the value must not be empty (\texttt{null}).
      \item[Level 2:] Business rules with additional logic, e.g., the ZIP code must consist of exactly four digits.
      \item[Level 3:] Logical combinations of rules. For example, to ensure the contact-ability of people either the e-mail address or phone number must be given (i.e., OR).
    \end{description}
\end{definition}

\begin{definition}[Minimality]
To what extent does a tool support duplicate detection?
    \begin{description}
      \item[Level 1:] Detection of exact duplicates on column-level. For example, two rows have the same social security number.
      \item[Level 2:] Detection of exact duplicates on row-level. For example, two rows have identical first names, last names, and social security number.
      \item[Level 3:] Similarity-based detection of duplicates. For example, two rows have identical first names and social security numbers, but due to a spelling mistake the last name differs.
    \end{description}
\end{definition}

\begin{definition}[Aggregation of Metrics]
To what extent can aggregations be part of a rule?
    \begin{description}
      \item[Level 1:] Aggregation of individual rule results. For example, $n$ DQ measures are aggregated into a single metric.
      \item[Level 2:] Manual weighting of rules. For example, rule A is only applied to a fraction of the data volume compared to rule B. When aggregating into a metric, the result of rule B is proportionally weighted higher than that of rule A.
      \item[Level 3:] Dynamic weighting of rules. For example, rule A is only applied to a fraction of the data volume compared to rule B. When aggregating into a metric, the result of rule B is automatically weighted higher than that of rule A.
    \end{description}
\end{definition}

\begin{definition}[Uncertainty]
To what extent do tools provide an indicator for confidence or uncertainty of individual rule results or aggregated metrics?
    \begin{description}
      \item[Level 1:] Uncertainty for duplicate detection. For example, the tool provides an additional indicator that quantifies the confidence or uncertainty of deduplication.
      \item[Level 2:] Standard uncertainty calculation for all rules. For example, the tool provides an additional indicator that quantifies the confidence or uncertainty of rule results.
      \item[Level 3:] Customizable uncertainty calculation per rule. For example, the tool provides a customizable indicator that quantifies the confidence or uncertainty of rule results.
    \end{description}
\end{definition}

\paragraph{Evaluation.}
\label{sec:evaluation}
The tabular overview in Table~\ref{tab:criteria_levels} provides a basis for comparing the tools with respect to the specified criteria. The evaluation is based exclusively on publicly available information, including vendor websites, accessible product documentation, and, where applicable, published GitHub repositories. ``NO'' / ``--'' indicates that the tool does not meet the criterion, while ``YES'' / ``+'' indicates that it does. For criteria with multiple levels, tools can achieve the levels ``+'' (Level 1), ``++'' (Levels 1 and 2), or ``+++'' (Levels 1, 2, and 3). The fulfillment of the criteria is assessed based on the defined examples.

\section{Evaluation Results}
\label{sec:results}

Table~\ref{tab:results} summarizes the evaluation results across all six tools and criteria. The following subsections discuss the findings in detail.

\begin{table*}[ht]
    \centering
    \caption{Tool evaluation based on the criteria defined in Section~\ref{sec:criteria}.}
    \begin{tabular}{l@{\hspace{15pt}}c@{\hspace{15pt}}c@{\hspace{15pt}}c@{\hspace{15pt}}c@{\hspace{15pt}}c@{\hspace{15pt}}c@{\hspace{15pt}}c}
        \textbf{Criteria} & \rotatebox{90}{\textbf{Great Expectations}} & \rotatebox{90}{\textbf{Deequ}} & \rotatebox{90}{\textbf{Evidently}} & \rotatebox{90}{\textbf{Informatica}} & \rotatebox{90}{\textbf{Experian}} & \rotatebox{90}{\textbf{Ataccama}} \\
        \toprule
        Open Source & Yes & Yes & Yes & No & No & No \\
        Quality Check with LLMs & No & No & No & No & No & No \\
        Rule Creation with LLMs & - & - & - & +++ & - & +++ \\
        Rule Definition (Correctness) & ++ & +++ & ++ & +++ & +++ & +++ \\
        Minimality & ++ & ++ & ++ & +++ & +++ & +++ \\
        Aggregation of Metrics & - & - & - & + & ++ & + \\
        Uncertainty & - & - & - & + & + & + \\
        \bottomrule
    \end{tabular}
    \label{tab:results}
\end{table*}

Generally, proprietary tools offer more comprehensive functionality and greater flexibility in defining rules for DQ checks, which are sometimes also supported by LLMs. These tools are often closed systems, in which functionality is divided into modules that must be purchased separately. This modality usually leads to vendor lock-in. Open-source tools are generally leaner, easier to integrate into existing tool ecosystems, and offer higher flexibility in terms of customization. However, they require significantly more effort for implementation and maintenance.

\subsection{Data Quality Checks with LLMs}
Although automated support for creating DQ metrics exists in nearly all tools, open-source tools rely on ML models or heuristic approaches, while proprietary DQ tools increasingly use LLMs due to their extensive systems. Nevertheless, existing tool providers do not use LLMs to validate the data itself, for example, to detect semantic data errors that are hard to detect for simple rules. However, there are initial approaches that use LLM assistants to detect anomalies in data profiling results.

\subsection{Rule-based Capabilities}
All examined tools support rule definition, as also previously found in~\cite{Ehrlinger_2022}. Proprietary tools additionally offer built-in functionality to combine multiple rules using logical operators. In most open-source tools, this functionality must be implemented manually.

Basic support for ensuring minimality is provided by all tools. Proprietary systems go a step further by offering duplicate detection based on similarity measurements to identify potential duplicates. Open-source tools typically rely on simple mechanisms to detect exact duplicates.

\subsection{Data Quality Metrics and Aggregation}
In the area of metric aggregation, both open-source and proprietary tools show significant limitations. Open-source tools do not offer direct support for aggregating numerical metrics. Instead, rules can be grouped and executed together. The result of such grouping shows the number of successful or failed rules but does not allow true numerical aggregation. Proprietary tools also offer only limited flexibility in terms of aggregations. Aggregations are usually predefined and restricted to specific data granularities. None of the tools examined allow the aggregation method to be freely configured (e.g., sum, average, median).

\subsection{Uncertainty}
An uncertainty indicator as a supplementary assessment for rule results is not offered by any of the open-source tools. In proprietary DQ tools, such functionality is only available in the context of deduplication. 
Tools such as Informatica and Experian provide categorical assessments that enable a rough evaluation of the quality of the results, whereas fine-grained or quantitative uncertainty evaluation is not performed.
\section{Conclusion}
\label{sec:conclusion}
Our evaluation reveals a clear distinction between open-source and proprietary DQ tools. Open-source tools such as Great Expectations, Deequ, and Evidently offer flexibility and ease of integration, but require significant implementation effort and lack advanced features such as uncertainty quantification and automated metric aggregation. Proprietary platforms such as Informatica, Experian, and Ataccama offer more comprehensive functionality and sophisticated rule management, at the cost of vendor lock-in and modular pricing.

Regarding LLM integration, a clear gap exists between research and practice. LLMs are increasingly used in research for DQ-related tasks such as error detection and data cleaning~\cite{glock2025,Ni_2024}, and vendors claim broad adoption of generative AI on their websites~\cite{gartner2021magicquadrant}. In practice, however, only two of the six evaluated tools, Informatica and Ataccama, integrate LLMs, and solely for rule creation. No tool currently uses LLMs for direct data validation, for example to detect semantic errors that are difficult to capture with simple rules. This represents a promising direction for future research and tool development~\cite{glock2025,Ni_2024}.

Beyond LLM integration, uncertainty handling and dynamic metric aggregation remain underdeveloped across all evaluated tools. None of the tools support freely configurable aggregation methods or fine-grained uncertainty quantification for rule results, which indicates clear potential for future innovation.

\begin{credits}
\subsubsection{\ackname} The research reported in this paper has been partly funded by the Federal Ministry for Innovation, Mobility and Infrastructure (BMIMI), the Federal Ministry for Economy, Energy and Tourism (BMWET), and the State of Upper Austria in the frame of the SCCH competence center INTEGRATE [(FFG grant no. 892418)] in the COMET - Competence Centers for Excellent Technologies Programme managed by Austrian Research Promotion Agency FFG and by the ``ICT of the Future'' project QuanTD (no. 898626).

\end{credits}
%
%
%
\bibliographystyle{splncs04}
\bibliography{bibliography.bib}

\end{document}